# STRONGLY NONLINEAR WAVES IN A CHAIN OF TEFLON BEADS

C. Daraio[1], V. Nesterenko[1,2*], E. Herbold[2], S. Jin[1,2]

*Materials Science and Engineering Program*[1]

*Department of Mechanical and Aerospace Engineering*[2]

*University of California at San Diego, La Jolla CA 92093-0411 USA*

**Abstract.** One dimensional "sonic vacuum" type phononic crystals were assembled from a chain of polytetrafluoroethylene (PTFE,Teflon) spheres with different diameters in a Teflon holder. It was demonstrated for the first time that this polymer-based "sonic vacuum", with exceptionally low elastic modulus of particles, supports propagation of strongly nonlinear solitary waves with a very low speed. These solitary waves can be described using classical nonlinear Hertz law despite the viscoelastic nature of the polymer and high strain rate deformation of the contact area. The experimentally measured speed of solitary waves at high amplitudes are close to the theoretically estimated values with a Young's modulus of 1.46 GPa obtained from shock wave experiments. This is significantly higher than the Young's modulus of PTFE from ultrasonic measurements. Trains of strongly nonlinear solitary waves excited by an impact were investigated experimentally and were found to be in reasonable agreement with numerical calculations based on Hertz interaction law though exhibiting a significant dissipation.

# I. INTRODUCTION

The study of strongly nonlinear wave propagation in one-dimensional chains of spherical beads representing the simplest model of granular materials has received much attention in recent years [1-32]. One of the distinguished properties of these materials is the existence of a qualitatively new solitary wave with a finite width that is independent of solitary amplitude. This solitary wave was first discovered in 1983 analytically and numerically [1] and later in 1985 it was observed in experiments [2]. *Strongly* nonlinear wave dynamics is a new area of interest, which is a natural extension of *weakly* nonlinear wave dynamics described by the Korteweg-de Vries equation [4,18]. Initially "strongly" precompressed, strongly nonlinear granular chains may behave as weakly nonlinear systems similar to the one considered in the Fermi-Pasta-Ulam paper [33]. In contrast to weakly nonlinear systems, the behavior of strongly nonlinear uncompressed granular chains exhibit qualitatively new features [4,17, 18, 23, 27, 30-32]. Novel applications might arise from understanding the basic physics of these 1-D systems. Sound focusing devices (tunable acoustic lenses and delay lines), sound absorption layers and sound scramblers are among the most promising engineering applications.

Non-classical, strongly nonlinear wave behavior appears if the granular material is "weakly" compressed [1,2,18]. In this case, the amplitude in a wave is significantly higher than the forces caused by initial precompression. The anharmonic approximation based on the small parameter (ratio of wave amplitude to initial precompression) is not valid. The principal difference between the strongly nonlinear case and the "strongly" compressed weakly nonlinear chain is due to the lack of a small parameter with respect to the wave amplitude in the former case. Long wave equation for particle displacement $u$ in this case is [1,18]:

$$u_{tt} = -c^2 \left\{ (-u_x)^{3/2} + \frac{a^2}{10} \left[ (-u_x)^{1/4} \left( (-u_x)^{5/4} \right)_{xx} \right] \right\}_x , \quad (1)$$

where,

$$-u_x > 0, \quad c^2 = \frac{2E}{\pi\rho(1-v^2)}, \quad c_0 = \left(\frac{3}{2}\right)^{1/2} c\xi_0^{1/4} .$$

Here $E$, $\rho$, and $v$ are the bulk elastic modulus, density, and Poisson ratio of the particles in the chain. The particle diameter is $a$ and $\xi_0$ is the initial strain in the system (phononic crystal). It should be mentioned that constant $c$ is the same order of magnitude as the bulk sound speed in the particle material and not the sound speed in the phononic crystal. Instead parameter $c_0$ corresponds to a long wave sound speed related to initial strain $\xi_0$. This equation for high amplitude pulses (or for negligible precompression) has no characteristic wave speed that is independent on amplitude. The regularized equation and the equation for a general interaction law can be found in [18]. Despite its complex nature Equation (1) has simple stationary solutions with unique properties that are similar to the stationary solutions for the discrete chain even though some differences due to the relatively short width of solitary wave exist [1,16,17,18,23,24,29]. The solitary wave speed $V_s$ in a "sonic vacuum" can be closely approximated by one hump of a periodic solution with finite length ($L$) equal only five particle diameters [1,18]:

$$\xi = \left(\frac{5V_s^2}{4c^2}\right)\cos^4\left(\frac{\sqrt{10}}{5a}x\right) = \frac{1}{8}\left(\frac{5V_s^2}{4c^2}\right)^2\left[3 + \cos 4\left(\frac{\sqrt{10}}{5a}x\right) + 4\cos 2\left(\frac{\sqrt{10}}{5a}x\right)\right]. \quad (2)$$

The concept of "sonic vacuum" was introduced in [3,5,6,9] to emphasize the fact that in an uncompressed chain ($\xi_0=0$) sound speed is equal to zero. The solitary wave speed $V_s$ has a nonlinear dependence on maximum strain $\xi_m$, the particle velocity $\upsilon_m$ and the force between particles $F_m$:

$$V_s = \frac{2}{\sqrt{5}}c\xi_m^{1/4} = \left(\frac{16}{25}\right)^{1/5} c^{4/5}\upsilon_m^{1/5} = 0.68\left(\frac{2E}{a\rho^{3/2}(1-\nu^2)}\right)^{1/3} F_m^{1/6}. \quad (3)$$

In a weakly compressed chain a supersonic solitary wave ($V_s > c_0$) with an amplitude much higher than the initial precompression propagates with a speed $V_s$, which can also be closely approximated by one hump of the periodic solution corresponding to zero prestress [1,18].

The speed of a wave in a "sonic vacuum" can be infinitesimally small if the amplitude of the wave is also small. It is interesting that a strongly nonlinear system supports solitary waves that are composed from a constant strain and only two harmonics (with wave length about 2.5$a$ and 5$a$ correspondingly) (see Eq. 2). The existence of this unique wave was verified analytically, numerically and in experiments [1,2,4,7,8,10,11,13,14,16,17,18,24]. This solitary wave can be considered as a soliton in a physically reasonable approximation [1,4,18], though small amplitude secondary solitary waves were observed in numerical calculations after collision of two identical solitary waves. The ratio of the largest amplitude of the secondary wave to the amplitude of the original wave is about 0.02 [21]. This solitary wave is of a fundamental interest because Equation (1) is more general than the weakly nonlinear KdV equation, which describes behavior of various physical systems [12] and the former includes the latter.

The solitary wave speed $V_s$ in a chain with finite prestress $\xi_0$ due to applied static precompression for tuning can be written in terms of normalized maximum strain $\xi_r=\xi_m/\xi_0$ or force $f_r=F_m/F_0$ [18]:

$$V_s = c_0 \frac{1}{(\xi_r - 1)} \left\{ \frac{4}{15} \left[ 3 + 2\xi_r^{5/2} - 5\xi_r \right] \right\}^{1/2} = 0.9314 \left( \frac{4E^2 F_0}{a^2 \rho^3 (1-\nu^2)^2} \right)^{1/6} \frac{1}{\left( f_r^{2/3} - 1 \right)} \left\{ \frac{4}{15} \left[ 3 + 2f_r^{5/3} - 5f_r^{2/3} \right] \right\}^{1/2}. \quad (4)$$

It is important to mention that $V_s$ can be significantly smaller than the bulk sound speed in the material composing the beads and can be considered approximately constant at any narrow interval of its relative amplitude $f_r$. The described properties of strongly nonlinear waves might allow the use of "sonic vacuum" based materials as effective delay lines with exceptionally low speed of signal propagation. The estimation based on Equation (3) with Young's modulus E=600 MPa, Poisson's ratio ν=0.46 and density ρ=2.2·10³ kg/m³ [34] of PTFE shows that it is possible to create materials with an impulse speed below 100 m/s, which corresponds to a particle velocity of 0.2 m/s or smaller. This signal speed in condensed soft matter is below the level of sound speed in gases at normal conditions. In this paper we present experimental results on pulse propagation in PTFE chains of spheres in accord with the main conclusions of the outlined strongly nonlinear theory. The speed of the signals is in the range of 88 m/s to 168 m/s. Uniformly compressed discrete chains are considered in numerical analysis and in experiments [4,18,25]. It was shown that the solitary wave speed generated by an impact of a piston with the same velocity increases with precompression. Also, the tendency of the impulse to split into a train of solitary waves decreases and the solitary wave width increases. Gravitationally loaded discrete chains are considered in numerical calculations in papers [15,19,22].

It should be noticed that particles with Hertzian contacts serving as strongly nonlinear springs are not the only way of discovering a "sonic vacuum" type system. Any power law interaction between particles (n>1) results in a similar behavior [3,17,18]. Also, any general strongly nonlinear interaction laws support solitary waves with finite length for the long wave approximation [18]. Different physical systems can be designed with properties suitable for the realization of "sonic vacuum" type behavior. For example, a forest of vertically aligned carbon nanotubes exhibits strongly nonlinear but non Hertzian type force interaction with spherical particles which can be used for assembling strongly nonlinear phononic crystals [35].

Solitary wave width for general strongly nonlinear interaction law is proportional to the bead diameter (or distance between particles) and the speed has a nonlinear dependence on amplitude [18]. It is interesting that a power law interaction with n=3, corresponding to a physical system of particles on an unstretched string in transverse vibrations [5,18] supports periodic harmonic waves and solitary waves with a linear dependence of maximum strain on speed.

## II. EXPERIMENTAL PROCEDURES AND RESULTS

One dimensional phononic crystals were assembled filling a PTFE tube (with inner diameter 5 mm) with chains of 11 and 21 PTFE balls (McMaster-Carr catalogue) with diameter $a$=4.76 mm and mass 0.1226 g (standard deviation 0.0008 g) (Fig. 1). Different numbers of particles were used to clarify the stages of impulse transformation and interaction with the wall. A chain assembled from 18 PTFE particles with smaller diameter $a$=2.38 mm and mass equal to 0.0157 g (standard deviation 0.0003 g) in a PTFE tube with inner diameter of 2.5 mm was also tested. Using two different sizes of beads helps to understand the behavior of the investigated polymeric material in the contact area at different stresses, strains and strain rate, which is dependent on the particle size. Scaling down the particle sizes is important for future applications in different devices (i.e. biomedical application, imaging, sound scrambling, etc.). Waves of different amplitude and durations were excited by impacting the top of the chain with strikers of different mass and velocities.

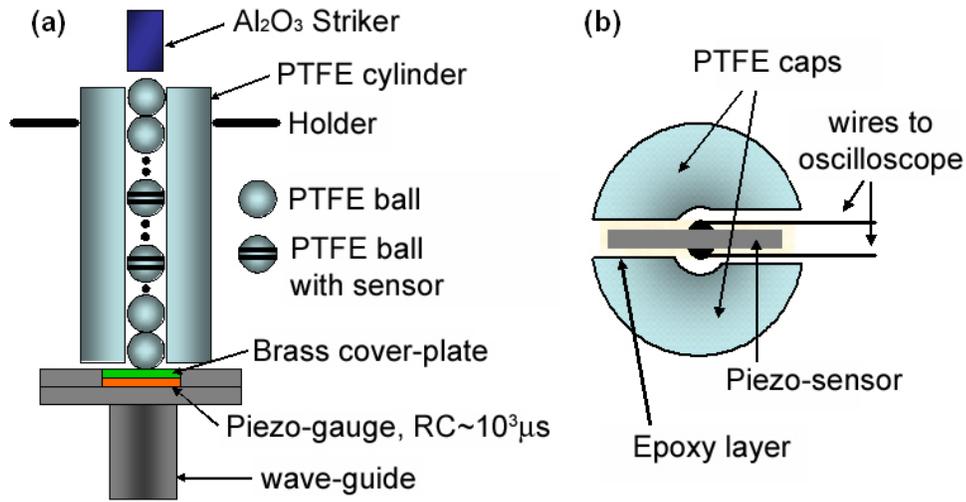

FIG. 1. (a) Experimental set-up for testing of 1-D strongly nonlinear phononic crystals with PTFE beads; (b) Schematic drawing of a particle with embedded piezo-sensor.

The experimental set-up for measurements of soliton speed, duration and force amplitude together with the measurements of reflected pulse from the wall is presented in Fig. 1. It includes three calibrated piezo-sensors (RC ~ $10^3$ μs) connected to a Tektronix oscilloscope. Two lead zirconate titanate based piezo-gauges (3 mm side plates with thickness 0.5 mm) with nickel plated electrodes and custom microminiature wiring, supplied by Piezo Systems, Inc., were embedded inside two PTFE particles similar to [18,36,37]. The particles with embedded sensors consisted of two PTFE caps with a total mass $2M$=0.093 g and a sensor with a mass $m$=0.023 g glued between these caps. Including glue, the total mass of the sensor was equal to 0.116 g (Fig. 1(b)), which was very close to the mass of the PTFE particle 0.123 g. This design allows a calculation of the speed of solitary wave simultaneously with measurement of the forces acting inside the particles.

A third piezo-gauge, supplied by Kinetic Ceramics, Inc., was bonded with epoxy on electrode pads for contacts and reinforced by a 1 mm brass plate on the top surface. The sensor assembly was then placed on the top surface of a long vertical steel rod (wave guide) embedded at the other end into a steel block to avoid possible wave reverberation in the system (Fig. 1(a)). This sensor was calibrated by using the impact of a single steel ball, which provides similar conditions of loading as in our measurements. Initial velocity and linear momentum conservation law were used for calibration. The area under the force-time curve measured by the gauge was integrated from the beginning of impact up to the point of maximum force and compared with linear momentum of particle at the beginning of impact. The sensors in the two particles were calibrated by comparison with the signal from the sensor at the wall. This was done using a controlled, relatively long, simultaneous loading of the particle with the sensor and the sensor in the wall by an impact of a massive piston.

The introduction of a particle with a different mass (particle with a sensor) in the chain of particle of equal masses results in wave reflections investigated in [11,17,19]. It was suggested to use reflected signals for detection of buried inclusions [19]. In numerical calculations, a slightly lighter particle with mass 0.116 g was introduced into the chain of particles with mass 0.123 g producing wave reflections that would be too small to detect experimentally. Attenuating soliton like pulse in a chain of random particles was considered in [1,4,18,20].

To interpret the signal measured in the experiments we considered the particle with an embedded sensor as a rigid body (Fig. 2(a)). The forces on the sides of the contacts of the particle ($F_1$ and $F_2$) can be easily related to the forces acting on both sides of the sensor ($F_3$ and $F_4$):

$$F_3 = \frac{F_1 + F_2}{2} + \frac{F_1 - F_2}{2} \frac{m}{2M + m}, \qquad F_4 = \frac{F_1 + F_2}{2} - \frac{F_1 - F_2}{2} \frac{m}{2M + m}. \qquad (5)$$

From Equation (5) we can see that the average of the compression forces, $F_3$ and $F_4$, (considered positive) is equal to the average value of forces $F_1$ and $F_2$, $F \equiv (F_1 + F_2)/2$ acting on the particle contacts. It should be mentioned that in numerical modeling the particles are considered rigid bodies and only contact forces $F_1$ and $F_2$ are taken into consideration. The time dependence of forces on the particle contacts were calculated numerically and their average values are presented in Fig. 2(b).

In the case of m<<M, the forces on each side of the sensor (Equation (5)) are very close to the average forces on the particle contacts [18,36]. In our case the forces $F_3$ and $F_4$ deviate from their average value by less than 20% in the vicinity of signal "shoulders", and are seen from the time dependence of contact forces in Fig. 2(b). It should be noted that at the moment when averaged force $F$ is maximum it is equal to the corresponding forces $F_3$ and $F_4$ (Fig. 2(b), Equation (5)). Comparison of averaged force and contact forces (Fig. 2(b)) reveals that averaging reduces the maximum amplitude of the force and increases the duration of the pulse. This averaged curve is used for comparison with experimental results based on sensors embedded in the particles.

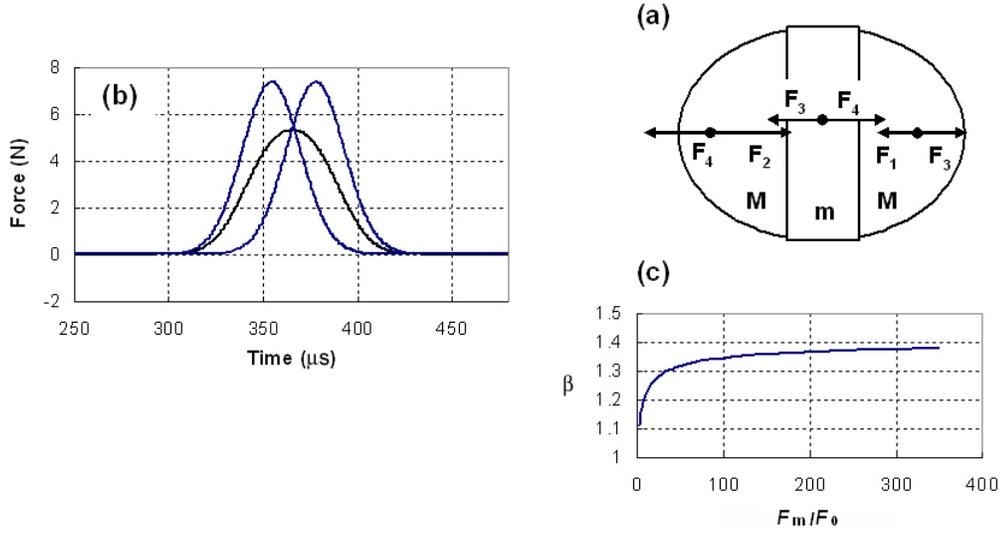

FIG. 2. (a) Schematic drawing showing the forces acting on the different parts of particle with embedded sensor; (b) Numerical calculation showing the forces vs. time obtained for the two contacts of the particle with embedded sensor (left and right curves correspondingly for $F_1$ and $F_2$) and average of the previous two (central curve) representing the average force acting on the sides of the sensor; (c) Dependence of the coefficient β on normalized solitary wave amplitude, $F_m/F_0$ with respect to static precompression.

To relate the maximum value of average compression force $F_{m.e}$ measured by the embedded sensor to the value of maximum force at the contact between neighboring particles (Fig. 2(a)), we used a coefficient β determined in numerical calculations. It represents the ratio of the dynamic force on the particles contacts to the maximum dynamic average of forces acting on the two contacts of the given particle in the solitary wave. Relating the two forces facilitates the use of Equations (3) and (4) with experimental data. The dependence of β on the relative force amplitude of a solitary wave is presented in Fig. 2 (c). This coefficient has negligible dependence on elastic moduli of PTFE particles (<1% in investigated range of a solitary wave amplitude and relevant range of elastic modulus from 600 MPa to 1460 MPa). It should be mentioned that the investigated range of relative amplitudes of dynamic force and static precompression represents a strongly nonlinear regime of system behavior resulting in relatively short length solitary waves. The coefficient β is about 1 in the linear regime when the amplitude of the dynamic force is much smaller than the initial precompression and the solitary waves are very long in comparison with a particle diameter.

The maximum compression force on the contact between two particles ($F_m$) adjacent to the sensor embedded in the particle was calculated using an equation:

$$F_m = \beta F_{m,e} + F_0, \qquad (6)$$

where $F_{m,e}$ is the maximum averaged dynamic compression force measured experimentally by the gauge embedded in the particle and $F_0$ is the gravitational precompression.

Pulses of different durations and amplitudes in the 1-D phononic crystals were generated by impact of an alumina ($Al_2O_3$) cylinder (0.47 g), a PTFE ball with a diameter of 4.76 mm (mass 0.123 g) or a stainless steel bead with a diameter of 2 mm (mass 0.036 g) onto the top particle of the chain. Single solitary waves can be generated by an impactor with a mass equal to the mass of the beads in the system which is physically equivalent to the application of δ−function force [1,4,18]. To generate a single solitary wave in a chain of 21 PTFE beads, we used the same bead as the striker (m=0.123 g). Sensors were placed in the 9[th] and 5[th] ball from the bottom and in the wall at the end of the chain.

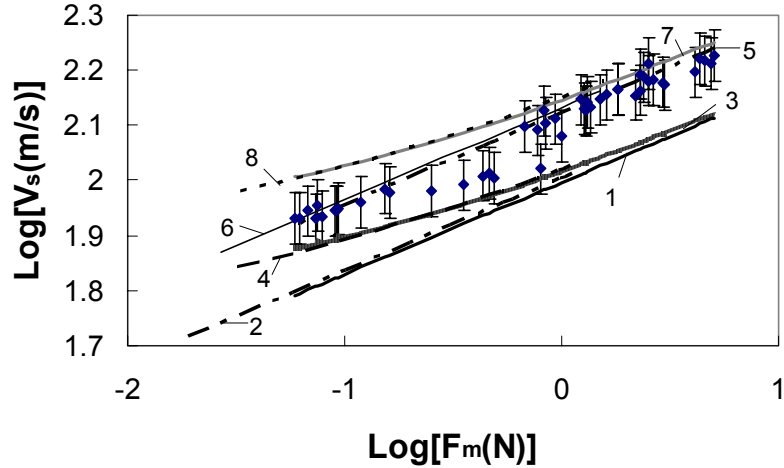

FIG. 3. Dependence of the velocity of solitary wave on amplitude. Experimental values are shown by solid dots. Curves 1 and 5 are the theoretical curves based on Equation (3) with a Young's modulus equal to 600 MPa and 1460 MPa respectively. Curve number 2 and 6 represent the corresponding numerical calculations for these cases. Curve number 3 and 7 represent long wave approximation for gravitationally pre-compressed systems (Equation (4)) at 600 MPa and 1460 MPa respectively; curves 4 and 8 represent the corresponding numerical calculation.

The theoretically predicted speed of solitary waves in strongly nonlinear phononic crystals has a strong dependence on the amplitude represented by Equations (3) and (4) for "sonic vacuums" and for precompressed chains respectively. This is shown in Fig. 3 together with the corresponding numerical calculations of the soliton speed for discrete chains. The curves based on the long wave approximation (Equations (3) and (4)) and the numerically calculated values practically coincide. In experiments, the solitary wave speeds for different amplitudes were obtained by dividing the distance between the sensors by the measured peak-to-peak time interval. The corresponding force amplitude in the solitary wave was found based on the measurements of gauges embedded inside the particles. The $\log F_m$-$\log V_s$ curves presented in Fig. 3 are based on these measurements.

Table 1. Experimental data for amplitude, speed, duration and normalized width of solitary wave in the PTFE chain composed of particles with diameter 2R=4.76 mm. Numerical data for discrete chains are also presented for comparison.

| | Experimental data | | | Numerical Results | | |
|---|---|---|---|---|---|---|
| $F_m$, [N] | Duration [μs] | $V_s$ [m/s] | $L/2R$ | $V_s$ [m/s] | $L/2R$ (at 0.2%) | $L/2R$ (at 4%) |
| 5 | 153 | 168 | 5.4 | 190 | 5.4 | 4.0 |
| 2 | 164 | 152 | 5.2 | 164 | 5.7 | 4.1 |
| 0.6 | 233 | 106 | 5.2 | 137 | 6.2 | 4.3 |
| 0.1 | 326 | 97 | 6.6 | 109 | 7.8 | 5.3 |
| 0.06 | 360 | 88 | 6.7 | 103 | 8.6 | 5.7 |

Accuracy of the measurements of amplitude of solitary waves was in the range of 15% to 30 % for large and small amplitudes. The larger errors being due to the higher signal to noise ratio at low amplitudes. In experiments the accuracy of the speed measurement can be estimated within 10 % due to the uncertainty in the sensors alignment (about 1 mm for each sensor).

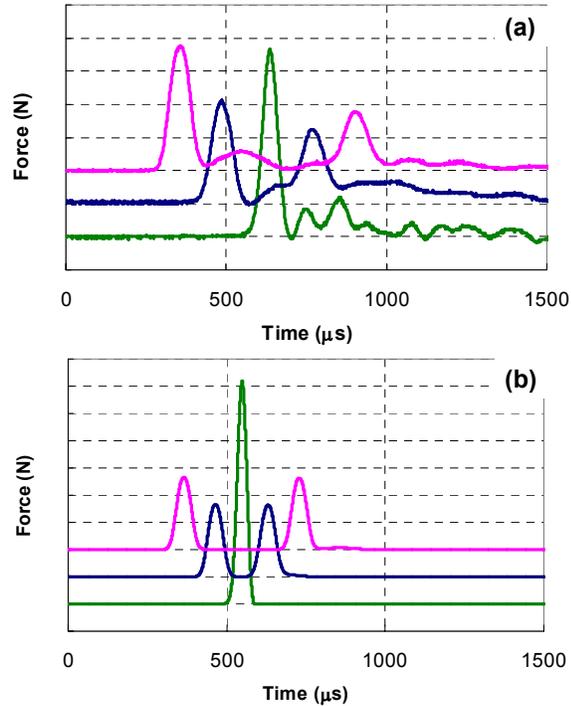

FIG. 4. Solitary waves in PTFE chain generated by PTFE ball-striker with a velocity of 2 m/s: (a) incident, reflected solitary waves and force on the wall detected experimentally in the chain of 21 PTFE beads with diameter 4.76 mm. The curves represent force vs. time detected by the sensor embedded into the 9$^{th}$ ball from the wall (top curve), and by the sensor in the 5$^{th}$ ball from the wall (middle curve) and at the wall (the vertical scale is equal to 0.5 N); (b) numerical calculations for a discrete chain under conditions corresponding to experimental conditions in (a). Curves represent the average value of the forces acting on the top and bottom contact of each sensor. Grid scaling on the vertical axes is 2 N.

After measuring the speed and duration of the propagating pulse, the widths of solitary waves were calculated for the corresponding force amplitudes (Table 1). The same data obtained from numerical analysis of discrete chains with PTFE elastic modulus 1.46 GPa based on the averaged forces on the particle contacts are also shown with solitary width truncated at the level 0.2% and 4% of the solitary wave amplitude.

Experimental results for forces measured by sensors embedded into the particles and into the wall corresponding to a 2.0 m/s impact velocity are shown in Fig. 4 (a). The zero time in all experiments corresponds to the start of recording triggered by the signal. In numerical calculations presented in all figures the zero time corresponds to the moment of impact.

One of the distinguished features of strongly nonlinear "sonic vacuum" type system is the fast decomposition of shock type pulse caused by impact on a short distance from the impacted side [1,4,18]. To check if the PTFE based strongly nonlinear phononic crystal exhibits this property, the impact by striker with a larger mass ($Al_2O_3$ cylinder 0.47 g) was employed to create a longer initial shock pulse in chains of different lengths. This impact results in an incoming pulse shape at the entrance of the system with the rise time equaling 50 μs and an initial decay with the characteristic exponential behavior $A\exp{-0.0185t}$, where the time ($t$) is measured in microseconds, starting from the peak of the signal with amplitude $A$. The total duration of the incident signal is equal to 370 μs. The result for the short chain composed of 11 PTFE particles is presented in Fig. 5 (a).

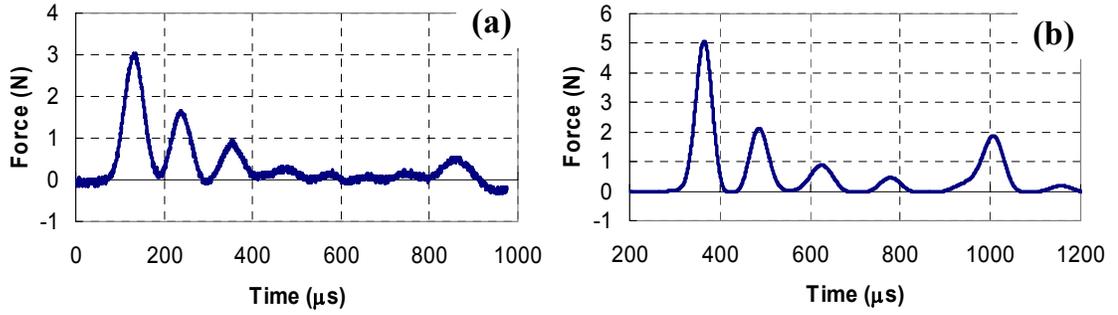

FIG. 5. Experimental and theoretical results demonstrating that a short chain of PTFE beads with a diameter 4.76 mm supports a train of solitary waves induced by the alumina striker with a mass equal four times the mass of the particle. (a) Force detected in experiment by the sensor mounted at the wall supporting an 11 PTFE particle chain, striker velocity 0.44 m/s, vertical scale is 1 N. (b) Numerical calculations corresponding to experimental conditions in (a), including gravitational precompression; vertical scale is 1 N, elastic modulus 1.46 GPa.

It is important to note that if exponential curves are drawn through the solitons maxima, corresponding to the force history at given point (in Fig. 5 this point corresponds to the wall), they will depend on the position of the sensor due to the dependence of soliton speeds on amplitude. Exponential decays corresponding to experimental data and numerical calculations at the wall are $A_e$exp-0.0067$t$ and $A_n$exp-0.0059$t$ (time is measured in microseconds). Absolute values of coefficients in these exponents are about three times smaller than in the incident pulse (0.0185) due to the dependence of soliton speeds on amplitude. Despite the evident attenuation in experiments ($A_e < A_n$) the corresponding exponents for envelope curves in experimental data and numerical calculations are close to each other. This suggests that the attenuation of the solitary waves is not strongly dependent on their amplitudes at the investigated range of amplitudes.

A chain of smaller diameter PTFE particles (2.38 mm) was also investigated to determine the diameter-dependence of strongly nonlinear behavior of PTFE-sphere-based "sonic vacuum" and dissipative properties. It should be mentioned that based on Hertz law the radius of the contact area is decreasing with particle radius under the same force. The experimental results are presented in Fig. 6 (a) for short duration of shock loading (impact by a 2 mm diameter steel ball with a mass about 2.3 times that of the PTFE particle) and for relatively long duration of impact induced by a PTFE ball with a diameter 4.76 mm and mass 0.123 g (Fig. 6 (c)).

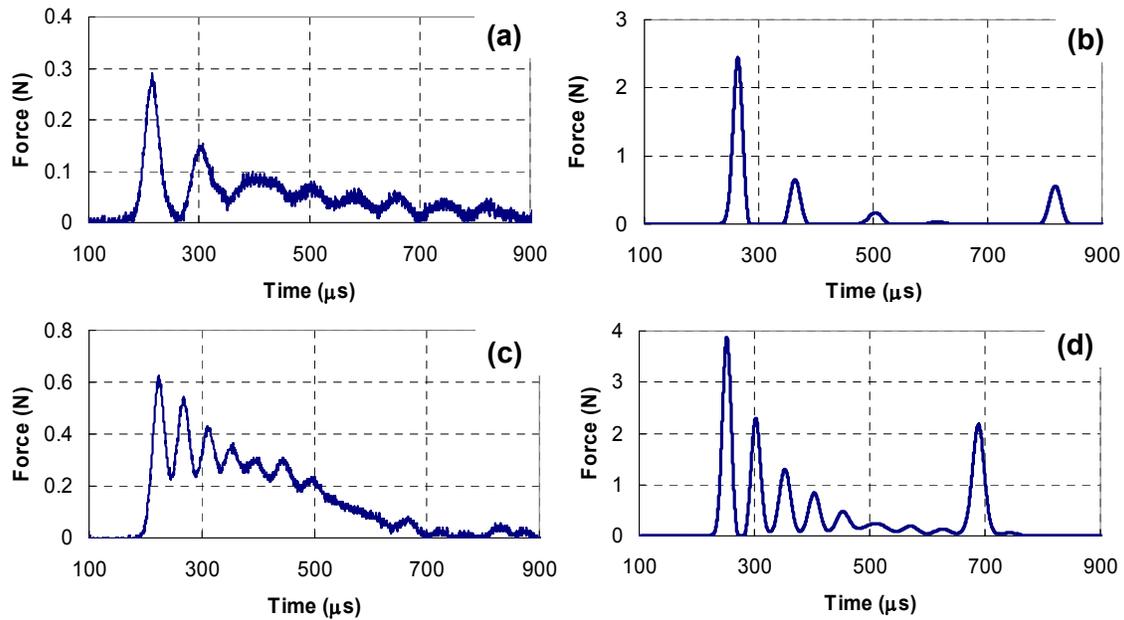

FIG. 6. The experimental and theoretical results demonstrating that a PTFE chain of smaller particles with 2.38 mm diameter supports the solitary waves and oscillatory "shock" waves modified by dissipation in experiments. (a) The leading solitary wave with an oscillatory tail detected at the wall generated in a chain of 18 smaller PTFE balls (0.016 g) (the velocity of 2 mm diameter steel ball impactor (0.036 g) was 0.89 m/s); (b) Numerical calculations corresponding to experimental conditions in (a), including gravitational precompression, elastic modulus 1.46 GPa; (c) Oscillatory "shock" wave detected at the wall generated in a chain of 18 PTFE balls, impacted at velocity 0.89 m/s with a 5 mm diameter PTFE ball (0.123 g); (d) Train of solitary waves detected in numerical calculations corresponding to experimental conditions in (c), including gravitational precompression, elastic modulus 1.46 GPa.

## III. NUMERICAL CALCULATIONS

It was shown previously [1,2,4,10,14,18] that wave propagation in 1-D system of linear elastic beads can be described considering particles as rigid bodies connected by nonlinear springs according to Hertz law (Eq. 7) for contact interaction of spheres [38,39].

$$F = \frac{2E}{3(1-\nu^2)}\left(\frac{R_1 R_2}{R_1 + R_2}\right)^{1/2}[(R_1+R_2)-(x_1-x_2)]^{3/2} \qquad (7)$$

A system of second order differential equations was reduced to the first order equations [1,4,18] and numerical calculations were performed using Matlab.

For comparison with experiments we calculated an average compression force $F=(F_1+F_2)/2$ for solitary waves based on the compression forces on the particle contacts ($F_1,F_2$). Both of these plots are shown in Figure 2(b). In our calculations we also used Hertz law for PTFE particle interaction. We used different values of elastic modulus and Poisson ratio equal to 0.46 [34,40]. Interaction between the flat wall and last particle was also described by Hertz law with an elastic modulus of 115 GPa and Poisson ratio of 0.307 for red brass (Cu85-Zn15). Hertz law was also used to calculate the interaction between the alumina impactor ($E$=416 GPa, $\nu$=0.231) and the first PTFE particle. No dissipation was included in numerical analysis. Gravitational force acting on particles causing initial nonuniform precompression in the chain increasing toward the wall was included in calculations. The linear momentum (before interaction with a wall) and energy were conserved with a relative error of $10^{-12}$% and $10^{-8}$%. Results of the numerical calculations modeling our experimental set-up and conditions are shown in Figs. 2 (b), (c), Fig.3, see corresponding curves, Fig. 4(b), Fig.5(b), and Fig.6(b), (d).

## IV. DISCUSSION

PTFE is a polymeric viscoelastic material with a strong strain rate sensitivity [41] and exceptionally low elastic modulus [40]. At normal conditions Young's and flexural moduli for PTFE are in the range of 400 – 750 MPa and Poisson's ratio is 0.46 [34,40,42]. This property can be very attractive for ensuring a very low speed of soliton propagation and tunability of the system. But it is not evident that a chain formed from this type of beads will support strongly nonlinear solitary waves as for chains made from typical linear elastic materials like stainless steel [2,10,14]. In particular, the role of dissipation and deviation from linear elastic law [41] for PTFE under high strain and high strain rate deformation in the contact area with high gradients of strain is the primary concern. In the present study, the strains obtained were up to 0.06 based on estimation of maximum compressive stresses in the center of contact on the order of 80 MPa at a maximum force approximately equal to 5 N. The typical strain rates were approximately $4 \times 10^2$ s$^{-1}$ and the compressive strains at the center of contact were about 0.06 decreasing to zero at a distance about 170 micrometers.

The beads made from Nylon with elastic modulus six times larger than that for PTFE demonstrated a Hertz type interaction law [14]. Chains made of these beads supported propagation of strongly nonlinear solitary waves with amplitudes in the range 1-33 N. In our experiments we extended range of amplitudes of solitary waves toward far smaller amplitudes by more than order of magnitude, down to 0.03 N. Dynamic behavior of "sonic vacuum" type systems at such low amplitudes is very interesting especially in view of potential practical applications related to noise reduction in audible acoustic range, acoustic lenses and delay lines, and for investigation of validity of Hertz law at very low displacements.

Furthermore, one of the distinguished features of "sonic vacuum" systems is a strongly nonlinear dependence of solitary wave speed on amplitude and precompression (Equations (3) and (4)) which are the important factors in imparting a tunability of various properties of these systems, for example in delay lines or acoustical lenses. Dependence of solitary wave speed on the amplitude ($F_m$), based on Equations (3) and (4), is shown in Fig. 3 (curves 1 and 3 respectively) with an elastic modulus equal to 600 MPa [43]. Numerical calculations based on the results for a discrete chain are also shown in Fig. 3 at the same value of elastic modulus (see curves 2 and 4). It is clear that theoretical and numerical approaches result in very close values of speeds at the given interval of amplitudes. For the low amplitude solitary waves, it was observed that gravitational precompression results in a noticeable deviation from the above curves. Comparison of the solitary wave profiles and speeds in a discrete chain and in a continuum approximation for different nonlinear interaction laws are considered in [17,29].

It is evident that there is a large difference between experimental values of soliton speed obtained in numerical calculations and in long wave theory for large amplitudes of force if the value of Young's modulus was taken as 600 MPa [43]. If 400 MPa is used for the PTFE elastic modulus [42], the difference between experimental speeds at high amplitude and predicted values based on the long wave approximation (or on numerical calculations) will be even more dramatic.

It should me mentioned that the dependence of shock wave speed $u_s$ on particle velocity $u_p$ in polymers (Hugoniots in $u_s(u_p)$ coordinates) extrapolated to bulk sound speed results in significantly higher values than the sound speed at normal conditions measured using ultrasonic technique. This well known discrepancy indicates a rapidly varying change of compressibility at low values of shock amplitudes [44]. For PTFE, the extrapolated value of bulk speed $c_b$ from Hugoniot gives a value of 1.68 km/s in comparison with 1.139 km/s from ultrasonic measurements. Using $c_b$=1.68 km/s from extrapolated Hugoniot measurements and Poisson ratio 0.46, we obtained a value of Young's modulus equal to 1.46 GPa based on relations for elastic solids [45]. Ultrasonic data for the same material gives a value of the elastic modulus equal to 704 MPa [45]. The calculated theoretical and numerical data for solitary wave speed versus amplitude using elastic modulus of 1.46 GPa are presented in the Fig. 3 (see curves 5-8). We can see that there is a better agreement between experimental data and calculated speed of solitary waves at high amplitude at 1.46 GPa and disagreement at low amplitudes.

Calculations with elastic modulus of 1000 MPa results in a reasonable correspondence between experimental data and calculation at the lower range of investigated force amplitudes of solitary waves. This suggests that the elastic modulus of PTFE is likely to be stress and strain rate dependent.

In a "sonic vacuum", solitary wave length does not depend on amplitude; it depends on the behavior of the interaction force [17,18]. In the case of a power law Hertz interaction ($n = 3/2$, Equation (6)) this length is equal to five particles. The properties of solitary waves were used to establish validity of Hertz law for different materials [14]. Measuring solitary wave speeds and durations in our experiments allows straightforward calculation of solitary wave widths corresponding to different amplitudes (see Table 1).

In the experiment corresponding to Fig. 4(a), for example, the speed of the leading solitary wave (with amplitude about 2 N) was measured using $5^{th}$ and $9^{th}$ particles from the wall and was found to be 152 m/s. The estimation based on the peak to peak measurements between the sensor in $5^{th}$ particle and the wall gave a similar value. The duration of solitary wave was about 164 microseconds resulting in the length of solitary wave equal to 5.2 times the diameter of the PTFE particle (Table 1). This is very close to the predicted length of solitary wave in long wave approximation equal to 5 particle diameters [1,4,18]. The measurements of averaged force using sensors embedded into the particle results in a slightly longer pulse in comparison with pulse duration based on the contact forces (Fig. 2(b)).

It is clear from the experimental data that the widths of solitary waves with relatively large amplitude are close to the predicted value of 5 particle diameter based on Hertz interaction law. Numerical analysis of a solitary wave in a "sonic vacuum" demonstrated that the energy contained in 5 particles is equal to 99.999996% of the total energy of the solitary wave. The distribution of velocities of particles in a solitary wave including more than 5 particles was analyzed in [16,46]. It should be mentioned that these widths do not depend on the elastic modulus. The widths of the solitary waves tend to be wider at lower amplitudes of propagating signals (Table 1). Numerical calculations performed in this work demonstrated a similar dependence of solitary widths with amplitude (see Table 1). This may be due to the influence of gravitational precompression.

It is evident from the comparison between Fig. 4(a) and Fig. 4(b) that numerical calculations of the behavior of a discrete system and experimental results are in close agreement with respect to the signal amplitudes and time durations between corresponding pulses. It is noted, however, that the amplitude of the reflected solitary wave recorded by the sensor inside the 9$^{th}$ bead is significantly smaller than the amplitude of the incident wave. This is apparently due to the presence of dissipation in experiments, which was not taken into account in numerical calculations and will be addressed in future research.

From the preceding discussions it is apparent that a chain of low modulus PTFE beads also supports the propagation of strongly nonlinear solitary wave, which is yet another realization of the "sonic vacuum" type phononic crystals with exceptionally low speed of signal.

Another remarkable feature of a "sonic vacuum" type system is the very fast decomposition of longer initial pulse into a train of solitary waves [1,2,4,17,18]. Apparently this phenomenon can be obscured by the strong dissipation in the system. To check if this property is also demonstrated by a chain of PTFE particles, we used a striker mass ($m_s$=0.47 g) that was higher than the mass of the particles in the chain to create a longer incident pulse.

Usually the number of solitary waves with significant amplitude is comparable to the ratio of the striker mass to the mass of the beads in the chain [1,4,17,18,28].

The results of this experiment are shown in Fig. 5. It is evident that this Teflon-based "sonic vacuum" also demonstrates very fast decomposition of initial impulse on a distance comparable with the soliton width (Fig. 5(a), (b)) and a clear tendency of signal splitting is very noticeable already after only 10 particles. The mass of the striker was chosen to be about four times that of the particles in the chain, expecting a decomposition of the initial triangular pulse into a train of four solitary waves. It should be mentioned that the number of solitary waves may be significantly larger if smaller amplitudes are included [17,46]. This example also demonstrates that a "short" duration impact on highly nonlinear "sonic vacuum" type ordered periodic systems results in a train of solitary waves instead of intuitively expected shock wave. An increase of the duration of impact results in a shock wave impulse with an oscillatory structure where the leading pulse can be of the KdV type solitary wave for weakly nonlinear chain or strongly nonlinear soliton with finite width for the strongly nonlinear case [1,4,18]**.** Similar qualitative agreement of the experimental results and numerical calculations was found for all investigated conditions of impacts.

Previous experimental work [2,7,8,10,14] with chains of steel beads, acrylic disks and spheres, glass, brass and nylon beads validated the prediction of strongly nonlinear solitary wave as stationary solutions of strongly nonlinear wave equation (Equation (1)). In those cases, the amplitude of the maximum force in the solitary wave was at least 3 times greater (1 N for Nylon beads) than the one obtained in this paper for PTFE beads (0.03 N). This and the higher elastic modulus of Nylon resulted in higher speeds of signal propagation (the minimum reported speed was 235 m/s for Nylon beads [14]). Furthermore, PTFE is a very versatile visco-elastic material. It is widely biocompatible, has a very low friction coefficient, and a very low elastic modulus which ensures applicability in a large variety of engineering solutions. As a result, we were able to experimentally achieve a speed of signal propagation of 88 m/s for a force amplitude of 0.06 N (Table 1), which is more than two times smaller than the speed of solitary wave detected for nylon beads [14] and more than 3 times smaller than the sound speed in the air at normal conditions. In principle, a "sonic vacuum" type media of different structure (Hertzian and non Hertzian) can support solitary waves with indefinitely small amplitude and speed of propagation. In the future, it is not unreasonable to expect that a "sonic vacuum" type system which supports detectable solitary waves with force amplitude similar to the one investigated in this paper with a speed of the order of magnitude of 10 m/s or lower could be designed using new materials with tailored elastic properties.

Finally, it is important to investigate the influence of particle size on the system behavior for application purposes. In fact, a smaller size of particles composing the PTFE based strongly nonlinear system results in different stresses and strain rate conditions in the contact area which may affect the system behavior. We conducted experiments with smaller diameter PTFE balls (2.38 mm) to check the validity of the strongly nonlinear theory. Experimental and numerical results are presented in Fig. 6. In the experiments, pulses were generated by impact of a 2 mm diameter steel ball (0.036 g, Fig. 6(a), (b) and a 5 mm diameter PTFE ball (0.123 g, Fig. 6 (c), (d)) at velocity 0.89 m/s. Numerical calculations did not account for the effects of dissipation. It is evident that the smaller diameter PTFE particles do support the "sonic vacuum" type behavior, although in this case the effect of dissipation appears to be more significant. Influence of dissipation on dynamics of solitary waves in strongly nonlinear discrete systems was considered in [20,26]. The effect of dissipation is likely to be responsible for the tail present after the 2nd solitary waves formed in experiments (Fig. 6(a)), and delays the solitary wave splitting in experiments in comparison with numerical results (compare Fig. 6 (a) to (b) and Fig.6 (c) to (d)).

## V. CONCLUSIONS

Propagation of impulses in one dimensional strongly nonlinear phononic crystals assembled from PTFE spheres was investigated for different conditions of loading and geometrical parameters. It was demonstrated, for the first time, that the chains of PTFE beads with different diameters support the Hertzian behavior with very low signal propagation speed due to its exceptionally low Young's modulus and despite viscoelastic nature of PTFE. Single solitary waves and decomposition of the signal into trains of solitary waves with amplitude more than one order of magnitude smaller than previously reported were observed. Small amplitude solitons broke the "sound barrier" having a speed of propagation well below sound speed in air. Single solitary waves and trains of strongly nonlinear solitary waves were excited by impact were investigated experimentally and were found to be in reasonable agreement with numerical calculations based on Hertz interaction law with Young modulus 1000 MPa for the lower amplitudes and 1460 MPa for higher amplitudes of signals, both being significantly higher than its value at normal conditions.

## ACKNOWLEDGEMENTS

This work was supported by NSF (Grant No. DCMS03013220).